\documentclass[pre,twocolumn,superscriptaddress,preprintnumbers,amsmath,amssymb]{revtex4-1}

\usepackage{epsfig}
\usepackage{amsmath}
\usepackage{graphicx}
\usepackage{fancybox}
\usepackage{subfigure}
\usepackage{color}
\usepackage{appendix}
\usepackage{sidecap}
\usepackage{comment}

\newcommand{\degree}{^{\circ} }
\newcommand{\erf}{\ensuremath{\mathop{\rm erf}}}
\newcommand{\erfc}{\ensuremath{\mathop{\rm erfc}}}
\newcommand{\rb}{\mathbf{r}}

\newcommand{\avg}[1]{\left<#1\right>}
\newcommand{\len}[1]{\left|#1\right|}
\newcommand{\brac}[1]{\left[#1\right]}

\newcommand{\para}[1]{\left(#1\right)}

\newcommand{\vect}[1]{\mathbf{#1}}
\newcommand{\gradr}{\nabla_{\rb}}

\newcommand{\vdiff}[2]{\left|\vect{#1} - \vect{#2}\right|}

\newcommand{\etal}{\emph{et al.}}

\newcommand{\rhoq}{\ensuremath{\rho^q}}

\newcommand{\rhob}{\rho^{\rm B}}

\newcommand{\phiRLJ}{\ensuremath{\phi_{\rm R}^{\rm LJ}}}
\newcommand{\phiRlLJ}{\ensuremath{\phi_{\rm R1}^{\rm LJ}}}

\newcommand{\Vrl}{\ensuremath{\mathcal{V}_{\rm R1}}}

\newcommand{\kT}{\ensuremath{k_{\rm B}T}}

\newcommand{\Fr}{\ensuremath{\mathcal{F}}}

\begin{document}


\title{Deconstructing classical water models at interfaces and in bulk}


\author{Richard C. Remsing}
\affiliation{Chemical Physics Program and Institute for Physical Science and Technology, University of Maryland, College Park, MD 20742}
\author{Jocelyn M. Rodgers}
 \affiliation{Physical Biosciences Division, Lawrence Berkeley National Laboratory, Berkeley, CA 94720}
\author{John D. Weeks}
\email{email: jdw@umd.edu}
\affiliation{Institute for Physical Science and Technology and Department of Chemistry and Biochemistry, University of Maryland, College Park, MD 20742}





\begin{abstract}
  Using concepts from perturbation and local molecular field theories
  of liquids we divide the potential of the SPC/E water model into
  short and long ranged parts. The short ranged parts define a minimal
  reference network model that captures very well the structure of the
  local hydrogen bond network in bulk water while ignoring effects of
  the remaining long ranged interactions. This deconstruction can
  provide insight into the different roles that the local hydrogen
  bond network, dispersion forces, and long ranged dipolar
  interactions play in determining a variety of properties of SPC/E
  and related classical models of water. Here we focus on the
  anomalous behavior of the internal pressure and the temperature
  dependence of the density of bulk water. We further utilize these
  short ranged models along with local molecular field theory to
  quantify the influence of these interactions on the structure of
  hydrophobic interfaces and the crossover from small to large scale
  hydration behavior. The implications of our findings for theories of
  hydrophobicity and possible refinements of classical water models
  are also discussed.
\end{abstract}
\maketitle

\section{Introduction}

Classical empirical water potentials involving fixed point charges and
Lennard-Jones (LJ) interactions were introduced in the first computer
simulations of water forty years ago and modern versions are widely
used even today in many biomolecular and materials-based
simulations. Two recent
reviews~\cite{GuillotJMolLiq2002,VegaIceFaraday2009} have focused on
this wide class of model potentials and assessed their performance for
a broad range of different structural and thermodynamic properties,
some of which were used as targets in the initial parameterization of
the models. Despite known limitations associated with the lack of
molecular flexibility and polarizability, they qualitatively and often
quantitatively capture a large number of properties of water and often
represent a useful compromise between physical realism and
computational tractability.

Given the simple functional forms of the intermolecular potentials it
may seem surprising that such good agreement is possible. But recent
work has shown that even simpler models where particles interact via
isotropic repulsive potentials with two distinct length scales are
able to qualitatively reproduce certain characteristic 
dynamic and thermodynamic
anomalies of bulk water~\cite{JaglaPRE,PRLTwoScalePots,JaglaSolvPNAS}.  Similarly in
dense uniform simple liquids a hard-sphere-like repulsive force
reference system can give a good description of the liquid structure,
and this in turn permits thermodynamic properties to be determined by
a simple perturbation theory~\cite{TheorySimpLiqs,WCA}.

This suggests it should be useful to analyze the construction and
predictions of empirical water potentials from the perspective of
perturbation theory of uniform fluids and the related Local Molecular
Field (LMF)
theory~\cite{LMFDeriv,LMFOGJPCB,LMFOGPNAS,LMFWater,WeeksAnnRevPhysChem,WCA}.
LMF theory provides a more general approach applicable to both uniform
and nonuniform fluids and gives strong support to the basic idea of
perturbation theory that in a uniform fluid slowly varying long ranged
parts of the intermolecular interactions have little effect on the
local liquid structure.

To apply these ideas to water we divide the intermolecular
interactions in a given water model into appropriately chosen short
and long ranged parts.  In this context, it is conceptually useful to
consider separately the slowly varying long ranged parts of both the
LJ interactions and the Coulomb interactions.  This deconstruction of
the water potential via LMF theory provides a hierarchical framework
for assessing separately the contributions of (i) strong short ranged
interactions leading to the local hydrogen bonding network, (ii)
dispersive attractions between water molecules, and (iii) long ranged
dipolar interactions between molecules.  Disentangling these
contributions without the insight of LMF theory is very difficult due
to the \emph{multiple} contributions of the point charges and the LJ
interactions in standard molecular water models

In uniform systems, the long ranged forces on a given water molecule
from more distant neighbors tend to cancel~\cite{WCA,WidomScience}.
The remaining strong short ranged forces between nearest neighbors
arise from the interplay between the repulsive LJ core forces and the
short ranged attractive Coulomb forces between donor and acceptor
charges.  These forces determine a minimal reference model that can
accurately describe the local liquid structure -- the hydrogen-bond
network for bulk water.  The slowly varying parts of the
intermolecular interactions are not important for this local structure
and could be varied essentially independently to help in the
determination of other properties as is implicitly done in the full
model.  Based on previous work with LMF
theory~\cite{LMFWater,MolPhysLMF,TrucCoul}, we examine two basic areas
where we expect the different contributions to play varying but
important roles -- bulk thermodynamics and nonuniform structure.  The
short ranged interactions responsible for the hydrogen-bonding network
are clearly necessary in all cases.  LMF theory allows us to determine
the relative importance of dispersive attractions and long-ranged
dipolar attractions in these applications using simple analytical
corrections for thermodynamics and an effective external field for
nonuniform structure.

In the next section we discuss the separation of the water potential
into short and long ranged parts and show that a minimal short ranged
reference model can very accurately describe atom-atom correlation
functions and other properties of the hydrogen bond network in bulk
water. In Sec.~3 we examine the thermodynamic implications of this
picture, focusing on two anomalous properties of bulk water
qualitatively well described by the full water model, the density
maximum at one atmosphere pressure and the behavior of the ``internal
pressure'' $(\partial E/\partial V)_{T}$, which has a temperature and
density dependence opposite to that of most organic solvents.  Then in
Sec.~4 we look at the effects of the unbalanced long-ranged
electrostatic and dispersion forces at aqueous interfaces. LMF theory
reveals strong similarities between the behavior of water at the
liquid-vapor interface and a planar hydrophobic wall, consistent with
previous work, and provides new insight into the relative importance
of electrostatic and dispersion forces and the transition from small
to large scale hydrophobicity as the radius of a nonpolar solute is
increased.

\begin{figure}
\centering
\includegraphics[width=0.4\textwidth]{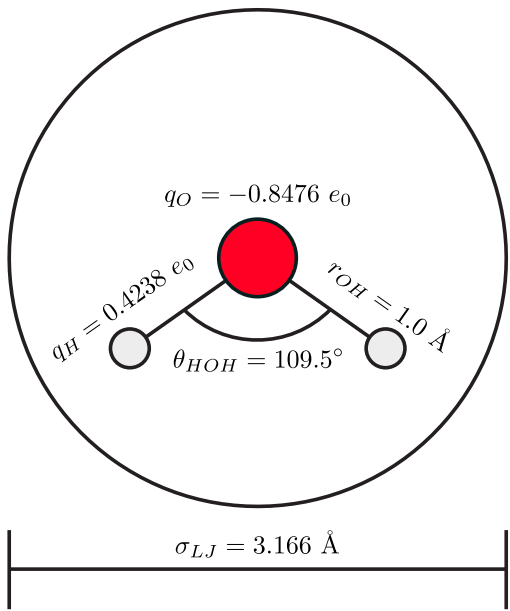}
\caption{(Color online) Schematic diagram of the SPC/E water model listing its
  various geometric parameters and interaction parameters. The O-H
  bond length and H-O-H angle are fixed, such that the molecule is
  rigid. The LJ well depth is $\epsilon_{LJ}=0.65$ kJ/mol. The oxygen
  site is depicted as a large red circle, while the hydrogen atoms are
  shown as smaller, gray circles.}
\label{fig:spce}
\end{figure}

\section{Local hydrogen bonds in full and truncated water potentials}

In this paper we consider one of the simplest and most widely used
water models, the extended simple point charge (SPC/E)
model~\cite{SPCE}, but similar ideas and conclusions apply immediately
to most other members of this class. As shown in Fig.~\ref{fig:spce}, SPC/E water
consists of a LJ potential as well as a negative point charge centered
at the oxygen site.  Positive point charges are fixed at hydrogen
sites displaced from the center at a distance of 1 \AA\ with a
tetrahedral HOH bond angle.  It is a remarkable fact that this simple
model can reproduce many structural, thermodynamic, and dielectric
properties of bulk water as well as those of water in nonuniform
environments around a variety of solutes and at the liquid-vapor
interface.

In the following we use the perspective of perturbation and LMF theory
to help us see how this comes about.  We use these ideas here not to
suggest more efficient simulations using short ranged model potentials
but rather as a method of analysis that provides physical insight into
features of the full model as well.  Since a detailed description and
justification of LMF theory is given elsewhere~\cite{LMFDeriv}, we
will focus on qualitative arguments and just quote specific results
when needed.

\begin{figure}
\centering
\includegraphics[width=0.5\textwidth]{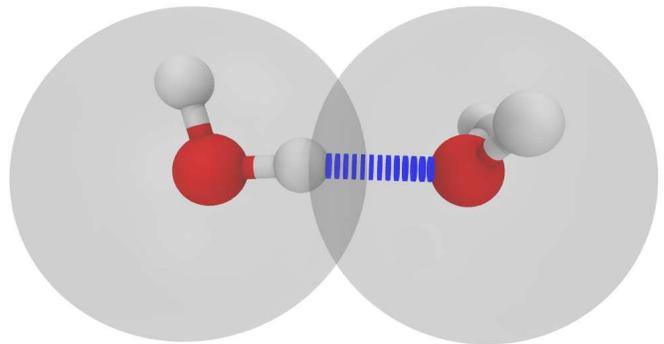}
\caption{(Color online) Optimal hydrogen bonding configuration of water taken from
  two molecules in ice Ih. LJ cores are depicted as gray transparent
  spheres with a diameter $\sigma_{LJ}=3.16$~\AA, while the hydrogen
  bond between waters with oxygens separated by 2.75~\AA\ is
  illustrated by a dashed, blue cylinder. Oxygen and hydrogen atoms
  are colored red and white, respectively.}
\label{fig:hbondoverlap}
\end{figure}

Fig.~\ref{fig:hbondoverlap} gives some insight into why a perturbation
picture based on the dominance of strong short ranged forces in
uniform environments could be especially accurate for bulk SPC/E and
related water models.  This shows two adjacent water molecules with a
separation of 2.75~\AA\ that form an optimal hydrogen bond as seen in
the structure of ice Ih.  Hydrogen bonding in this model is driven by
the very strong attractive force between opposite charges on the
hydrogen and oxygen sites of adjacent properly oriented molecules.
Proper orientation permits similar strong bonds to form with other
molecules, leading to a tetrahedral network in bulk water.  The gray
circles drawn to scale depict the repulsive LJ core size as described
by the usual parameter $\sigma_{LJ}=3.16$~\AA. The substantial overlap
indicates a large repulsive core force opposing the strong
electrostatic attraction, finally resulting in a nearest neighbor
maximum in the the equilibrium oxygen-oxygen correlation function of
2.75~\AA.

It is interesting to note that the first BNS water model introduced in
1970 used a smaller core size
$\sigma_{LJ}=2.82$~\AA~\cite{BNSModel-Book}.  However a much larger LJ
core with strong core overlap at typical hydrogen-bond distances is a
common property of almost every water model introduced since then and
seems to be a key feature needed to get generally accurate results
from simple classical point charge models.  Evidentially the highly
fluctuating local hydrogen-bond network in these models arises from
geometrically-frustrated ``charge pairing'', where the strong LJ core
repulsions and the presence of other charges on the acceptor water
molecule oppose the close approach of the strongly-coupled donor and
acceptor charges.

We can test the accuracy of this picture by considering various
truncated or ``short'' water models where slowly varying long ranged
parts of the Coulomb and LJ interactions in SPC/E water are completely
neglected.  We first consider a Gaussian-truncated (GT) water model,
already studied by LMF theory~\cite{LMFWater,MolPhysLMF,TrucCoul}.
Here the Coulomb potential is separated into short and long ranged
parts as
\begin{equation}
\label{eq:coulombsep}
v(r)=\frac{1}{r}=\frac{\erfc(r/\sigma)}{r}+\frac{\erf(r/\sigma)}{r}=v_0(r)+v_1(r),
\end{equation}
where $\erf$ and $\erfc$ are the error function and complementary
error function, respectively.  The short-ranged $v_0(r)$ is the
screened electrostatic potential resulting from a point charge
surrounded by a neutralizing Gaussian charge distribution of width
$\sigma$. Hence $v_0(r)$ vanishes rapidly at distances $r$ much
greater than $\sigma$ while at distances less than $\sigma$ the force
from $v_0(r)$ approaches that of the full $1/r$ potential.

\begin{figure}
\centering
\includegraphics[width=0.5\textwidth]{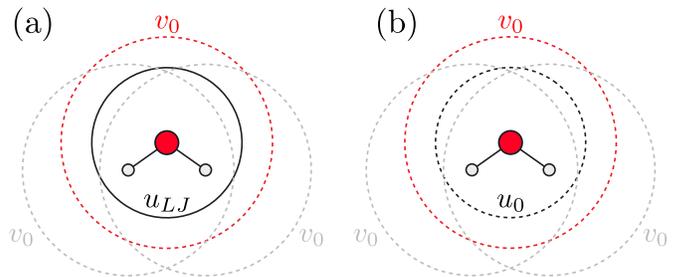}
\caption{(Color online) Schematic diagrams of (a) GT and (b) GTRC water
  models. Truncated interactions are indicated by dashed lines, while
  full interaction potentials are indicated by solid lines. LJ
  interactions are represented by black lines, while oxygen and
  hydrogen electrostatic interaction potentials are shown as red and
  gray lines, respectively.}
\label{fig:sketch}
\end{figure}

In GT water, depicted in Fig.~\ref{fig:sketch}a, the Coulomb potential
associated with each charged site in SPC/E water is replaced by the
short-ranged $v_0$ with no change in the LJ interaction.  As suggested
by Fig.~\ref{fig:hbondoverlap}, important features of the local
hydrogen-bond network should be well captured by such a truncated
model if the cutoff distance controlled by the length parameter
$\sigma$ in Eq.~(\ref{eq:coulombsep}) is chosen larger than the
hydrogen bond distance.  Following Refs.~\cite{MolPhysLMF}
and~\cite{TrucCoul}, here we make a relatively conservative choice of
$\sigma=4.5$~\AA, but values as small as 3~\AA\ give essentially the
same results. The circles are drawn to scale with diameters $\sigma$ and
$\sigma_{LJ}$.

The basic competition between very strong short ranged repulsive and
attractive forces in the hydrogen bond depicted in
Fig.~\ref{fig:hbondoverlap} should be captured nearly as well by an
even simpler reference model where the LJ potential is truncated as
well, and replaced by the repulsive force reference potential
$u_{0}(r)$ used in the WCA perturbation theory for the LJ
fluid~\cite{WCA}. The resulting Gaussian truncated repulsive core
(GTRC) model is schematically shown in Fig.~\ref{fig:sketch}b.

\begin{figure*}
\centering
\includegraphics[width=.98\textwidth]{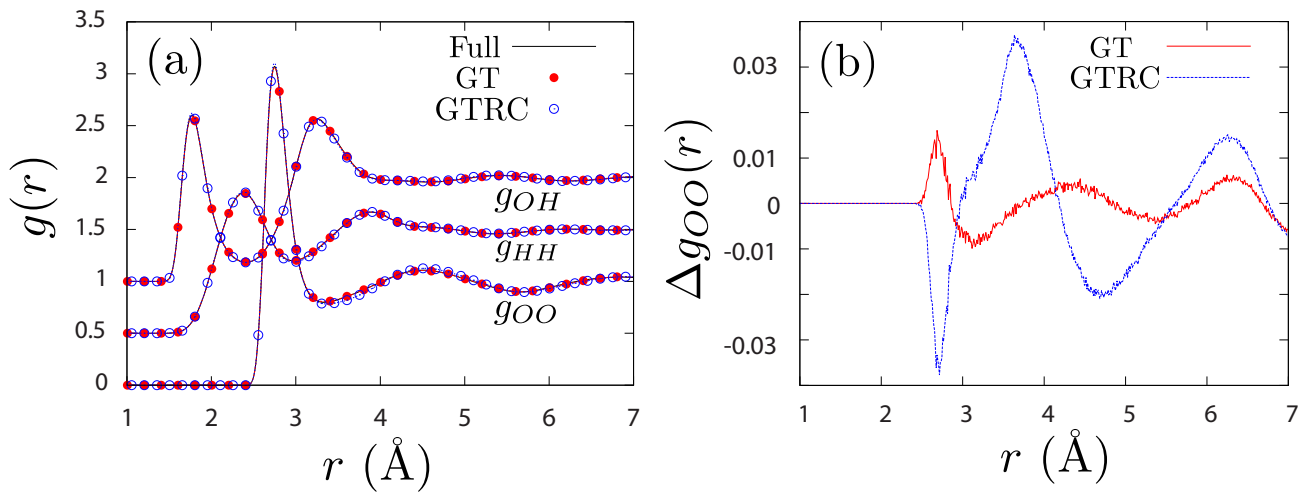}
\caption{(Color online) (a) Oxygen-oxygen, oxygen-hydrogen, and hydrogen-hydrogen
  site-site pair distribution functions, $g_{OO}(r)$, $g_{OH}(r)$, and
  $g_{HH}(r)$, respectively, for the three water models under study at
  $T=300$~K and $v=30.148$~\AA$^3$. $g_{HH}$ and $g_{OH}$ have been
  shifted by 0.5 and 1 units, respectively, for clarity. (b)
  Differences between $g_{OO}(r)$ of the full model and that of the
  designated reference systems, $\Delta g_{OO}(r)$.}
\label{fig:RDFs}
\end{figure*}

As discussed in perturbation theories of simple
liquids~\cite{TheorySimpLiqs,WCA}, a well-chosen reference system
should accurately reproduce bulk structure present in the full system
at the same fixed density and temperature. As illustrated by the pair
distribution functions in Fig.~\ref{fig:RDFs}, bulk GT and GTRC water
models have a liquid state structure virtually identical to that in
the full SPC/E model.  This very good agreement is also reflected in
other properties of the hydrogen-bond network. We directly examined
the hydrogen bonding capabilities of GT and GTRC water models through
the calculation of the average number of hydrogen bonds per water
molecules, $\avg{n_{HB}}$, as well as the probability distribution of
a water molecule taking part in $n_{HB}$ hydrogen bonds,
$P\para{n_{HB}}$, using a standard distance criterion of hydrogen
bonds, $R_{\rm OO}<3.5$~\AA\ and $\theta_{\rm HOO'}<30\degree$, where
$R_{\rm OO}$ is the oxygen-oxygen distance and $\theta_{\rm HOO'}$ is
the angle formed by the H-O bond vector on the hydrogen bond
\textit{donating} water molecule and the O-O$'$ vector between the
oxygen on the donor water (O) and the acceptor oxygen
(O$'$)~\cite{LuzarChandlerPRLHbondDef-1996}.  For both GT and GTRC
water models, $\avg{n_{HB}}$ and $P\para{n_{HB}}$ were calculated at
temperatures ranging from 220-300 K, and were found to be nearly
identical to the analogous quantities in the full SPC/E model (data
not shown).  These findings give credence to the idea that these two
truncated models reproduce the hydrogen-bond network of the full model
to a high degree of accuracy.
 
These truncated models offer a minimal structural representation of
bulk water as a fluctuating network of short ranged bonds determined
mainly by the balance between the very strong electrostatic attraction
between donor and acceptor charges and the very strong repulsion of
the overlapping LJ cores. We can view them as primitive water models
in their own right, analogous to other simplified models recently
proposed, which capture very well arguably the most important
structural feature of bulk water, the hydrogen bond network, and it is
instructive to see what other properties of water such minimal network
models can describe.  But corrections from neglected parts of the
intermolecular interactions are certainly needed for bulk
thermodynamic and dielectric properties and for both structure and
thermodynamics of water in nonuniform environments. LMF theory
provides a more general framework where the truncated models are
viewed as useful reference systems that can be systematically
corrected to achieve good agreement with full water models. We will
use both viewpoints in the next section.

\section{Thermodynamic anomalies of bulk water using full and truncated water potentials}

\subsection{Density maximum}
Now we turn our attention to the thermodynamics of bulk water. For a
fixed volume $V$, temperature $T$, and number of molecules $N$, the
pressure and other thermodynamic properties of the GT and GTRC systems
will not generally equal those of the full system.  However, because
of the accurate reference structure, we can correct the thermodynamics
using simple mean-field (MF) arguments. Thus we can define the
pressure in the full system to be the sum of the short-ranged
reference pressure and a long-ranged correction, $P=P_0+P_1$.

Simple corrections to the energy and pressure of the GT model from
this perspective were recently derived~\cite{TrucCoul}. With $\sigma
=4.5$~\AA, these corrections are relatively small and were ignored in
most earlier work using truncated water models but they are
conceptually important in revealing the connections between truncated
models and perturbation theory and are required for quantitative
agreement.  The pressure correction $P_1=P^q_1$ for the GT model
arises only from long-ranged Coulomb interactions and is given as
\begin{equation}
\label{eq:P1q}
P^q_1=-\frac{\kT}{2 \pi^{3/2}\sigma^3}\frac{\epsilon-1}{\epsilon},
\end{equation}
where $\epsilon$ is the dielectric constant.

In the case of the GTRC model, the need for a thermodynamic correction
is much more obvious since we have to correct for the absence of LJ
attractions as well. Here we adopt the simple analytic correction used
in the van der Waals (vdW) equation derived from WCA theory for the LJ
fluid, as discussed in Ref.~\cite{WeeksAnnRevPhysChem}. Thus
$P_1=P^q_1-a\rho^2$ for the GTRC potential, where
\begin{equation}
\label{eq:vdwa}
a\equiv -\frac{1}{2}\int d\mathbf{r}_{2}\,u_{1}(r_{12})
\end{equation}
and $u_{1}$ is attractive part of the LJ potential.  This simple
approximation does not give quantitative results but does capture the
main qualitative features and we use it here to emphasize the point
that both the long ranged Coulomb and dispersion force corrections to
bulk GTRC water can be treated by simple perturbation methods.

We can test the accuracy of these corrections by using them to help
determine the temperature $T_{MD}$ at which the density maximum of the
full SPC/E water model at a constant pressure of 1~atm should occur.
This can alternatively be defined as the temperature at which the
thermal expansion coefficient, $\alpha_P$, is zero. Accordingly, we
seek to evaluate $\alpha_P$ using the relation
\begin{equation}
  \alpha_P\equiv \frac{1}{v}\para{\frac{\partial v}{\partial T}}_{P}=-\frac{1}{v}\para{\frac{\partial P}{\partial T}}_v \para{\frac{\partial v}{\partial P}}_T,
\label{eq:thermexp}
\end{equation}
where $v=V/N$ is the volume per particle.  Using the last expression
we can determine where the quantity $\para{\partial P/\partial T}_v=0$
by monitoring the corrected pressure of the reference systems while
changing the temperature at a fixed density.  This can be done by
simulation in the canonical ensemble.  The fixed density ensures that
the structure of the reference and full systems are very similar, as
assumed in the derivation of the corrections in Eqs.~(\ref{eq:P1q})
and~(\ref{eq:vdwa}).  We can also determine $T_{MD}$ through the first
equality in Eq.~(\ref{eq:thermexp}) by finding where
$\para{\partial v/ \partial T}_P=0$.  Thus we simulate GT and GTRC
water at constant pressures of $P_0=P-P_1$, where $P=1.0$~atm is the
pressure in the full system. Note that the correction $P^q_1\equiv
P^q_1(T;\epsilon(T))$ is temperature-dependent, as is the dielectric
constant $\epsilon$, so that we are not moving along an isobar in
$P_0$, but an isobar in $P$.  The temperature-dependent values of
$\epsilon$ were taken to be the experimental values~\cite{CRC}.

\begin{figure}
\centering
\includegraphics[width=.5\textwidth]{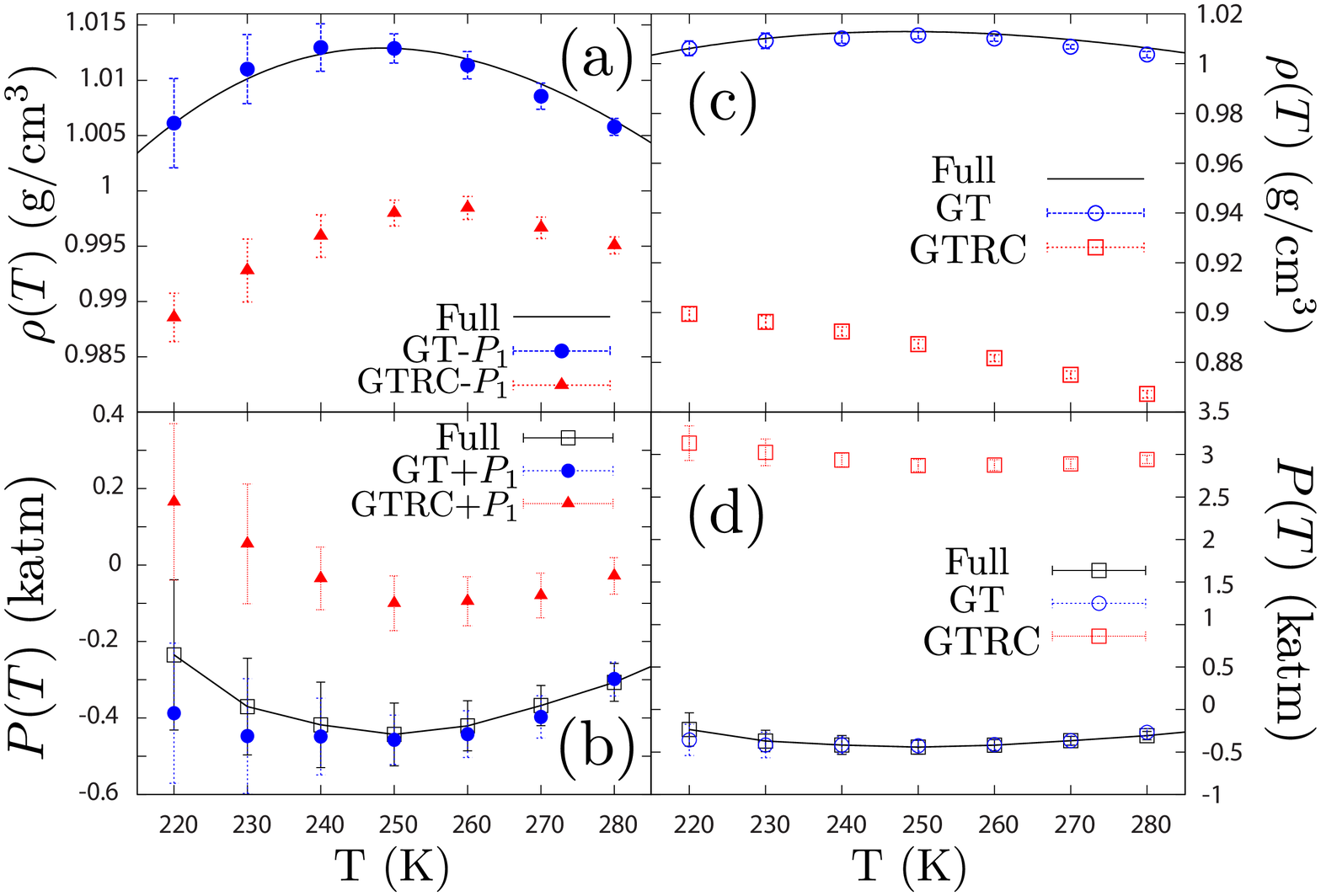}
\caption{(Color online) The dependence of (a) density and (b) pressure for the
  corrected reference models as a function of temperature. The
  analogous quantities for the models without MF corrections are shown
  in (c) and (d), respectively. $\rho(T)$ is calculated at a constant
  pressure of 1.0~atm and $P(T)$ is calculated at a fixed volume of
  $v=30.148$~\AA$^3$. Full SPC/E data for $\rho(T)$ at constant P was
  taken from the work of Ashbaugh \etal~\cite{AshbaughTMD}.}
\label{fig:density}
\end{figure}

Figs.~\ref{fig:density}a and~\ref{fig:density}b give the density and
pressure of full SPC/E water and the corrected reference models as a
function of temperature.  As expected, the inclusion of $P^q_1$ in the
pressure of GT water quantitatively corrects the density and pressure
of this system.  However, the MF correction applied to GTRC water,
$P_0=P-P^q_1+a\rho^2$, is not as accurate, although the dependence of
$\rho$ on $T$ is qualitatively well captured.  These remaining errors
arise from our use of the simple van der Waals $a\rho^2$ correction
for the long ranged part of the LJ potential.  This level of agreement
is typical when this correction is used in pure LJ
fluids~\cite{WeeksAnnRevPhysChem} and a full WCA perturbative
treatment of the attractive portion of the LJ potential in GTRC water
would likely lead to quantitatively accurate results~\cite{WCA}.

We now turn to the alternate and less accurate interpretation of the
GT and GTRC models as primitive water models in their own right. Do
these models at an uncorrected pressure of 1~atm have a density
maximum and how well does it compare to that of the full model? To
that end, we find where $\para{\partial v/\partial T}_P=0$ in each
model by varying the temperature along an isobar using MD simulations
in the isothermal-isobaric ensemble at a constant pressure of 1~atm.
By requiring the same pressure in the full and reference models, we
probe structurally different state points in general and there is no
guarantee that the density and temperature of the reference systems at
a density maxima (if present) will be similar to that in the full
system. Nevertheless Fig.~\ref{fig:density}c shows that the GT model
does have a density maximum very similar to that of the full model.
This is because the pressure correction to the density from the
long-ranged Coulomb interactions in Eq.~(\ref{eq:P1q}) is very small
on the scale of the graph.  In contrast, the uncorrected GTRC model
does not exhibit a density maximum at $P=1.0$~atm, even upon cooling
to 50 K (data not shown).

\begin{figure*}[t]
\centering
\includegraphics[width=.98\textwidth]{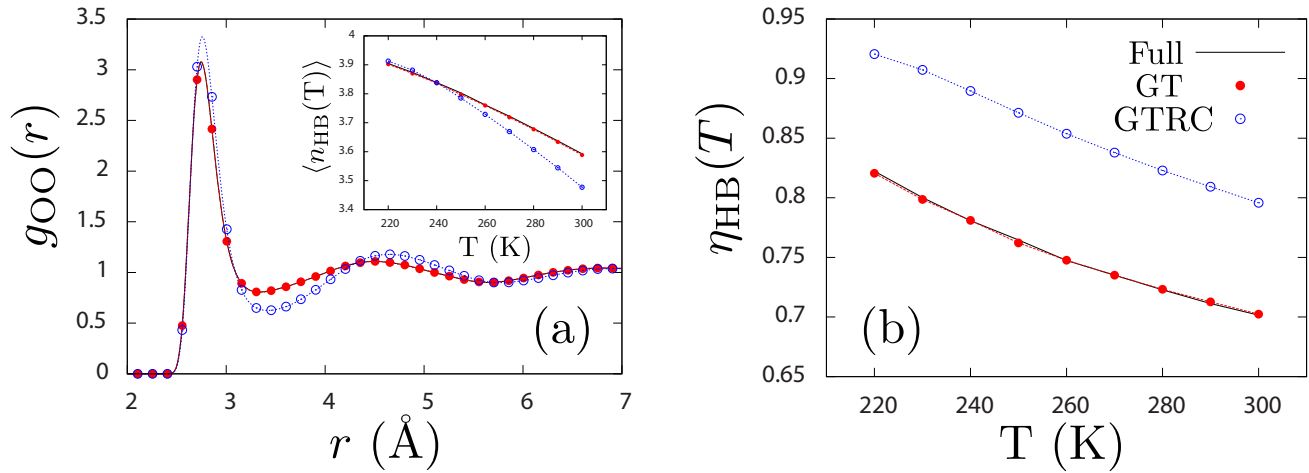}
\caption{(Color online) (a) The oxygen-oxygen pair distribution function, $g_{\rm
    OO}(r)$, for the three water models at $T=300$ K.  $Inset$: The
  average number of hydrogen bonds per water molecule as a function of
  temperature, $\avg{n_{\rm HB}(\rm T)}$, for full and truncated water
  models. (b) Hydrogen bonding efficiency $\eta_{\rm HB}$ as a
  function of temperature. All results were obtained at a constant
  uncorrected pressure of 1~atm.}
\label{fig:hbond}
\end{figure*}

These results should be compared to earlier work where the TIP4P water
potential was approximated by a simpler ``primitive
model''~\cite{NezbedaThermo}.  In that work, the repulsive LJ core was
mapped onto a hard-sphere potential, hydrogen bonding was captured by
a square-well potential, and long-ranged dipole-dipole interactions
were represented with a dipolar potential. The equation of state was
found using a perturbative approach, and thermodynamic quantities were
analyzed.  The authors of Ref.~\cite{NezbedaThermo} found that the
inclusion of dispersion forces does not lead to a density maximum, and
only when both dispersive interactions and long-ranged dipole-dipole
interactions were taken into account did a density maximum appear.

To provide some understanding of these differing results, we analyze
the structure of the uncorrected GT and GTRC reference models in
comparison to the full model at the common pressure of one
atmosphere. The oxygen-oxygen radial distribution functions, $g_{\rm
  OO}(r)$, for each of the three water models at $T=300$ K are
depicted in Fig.~\ref{fig:hbond}a. The GT model is in good agreement
with the full model, consistent with its accurate description of the
bulk water density and the density maximum.  In contrast, as shown
later in Fig.~\ref{fig:lvint}, the coexisting liquid density of GTRC
water is about about 15\% lower than that of the full water model.
Nevertheless the first peak of $g_{\rm OO}(r)$ in GTRC water is
\emph{higher} than that of the full water model due to better
formation of local hydrogen bonds. As shown in the inset, a molecule
of GTRC water has slightly fewer hydrogen bonds on average than full
and GT water models for temperatures higher than 240 K. However the
hydrogen bond efficiency shown in Fig.~\ref{fig:hbond}b,
\begin{equation}
\eta_{\rm HB}=\frac{\avg{n_{\rm HB}}}{\avg{n_{\rm NN}}},
\end{equation}
where $\avg{n_{\rm NN}}$ is the average number of nearest-neighbors
satisfying $R_{\rm OO}<3.5$~\AA, indicates that GTRC water is about 10
percent more efficiently hydrogen bonded to its available neighbors at
all temperatures. In this sense the low density GTRC water at
$P=1.0$~atm is structurally more ice-like than the full water model.

These results provide some insight into earlier
first principles simulations of liquid water 
using density functional theory~\cite{DFTNPTwater,DFTwaterLV, DFTvdWJCP}.
The standard exchange-correlation functionals used there can give a good description
of local hydrogen bonding, but do not include effects of van der Waals interactions.
These simulations produced a decrease in the bulk density of water accompanied by increased
local structural order very similar to that seen here for GTRC water. Moreover, when
dispersive interactions were crudely accounted for, they observed much better
agreement with experiment, in 
complete agreement with our findings for perturbation-corrected GTRC water.

Our results indicate that van der Waals attractions play the role of a
cohesive energy needed to achieve the high density present in SPC/E
water at low pressure, as demonstrated by the qualitative accuracy of
Eq.~(\ref{eq:vdwa}) and the good agreement of the GT model.
Evidentially a density maximum can arise only when additional somewhat
less favorably bonded molecules are incorporated into the GTRC network
to produce the full water density.  If the local hydrogen bond network
of water at the correct bulk density is properly described, long
ranged dipolar forces are not needed to obtain the correct behavior of
$\rho(T)$.  Indeed, LJ attractions are not needed either provided that
the proper bulk density is prescribed by some other means. Thus we found
that if GTRC water is kept at a high constant pressure of 3~katm,
where its bulk density is close to that of the full water model at ambient
conditions, a density maximum is also observed (data not shown).

\subsection{Internal pressure}
We further employ the reference water models to explain the anomalous
``internal pressure'' of water~\cite{PrattWaterQCT-JCP2007}.  For a
typical van der Waals liquid, the internal pressure is given by
$P_i=\para{\partial \varepsilon/\partial v}_T\approx a\rho^2$ for low
to moderate densities, where $\varepsilon=E/N$ is the energy per
molecule. In fact, it was recently shown by computer simulation that
the portion of the internal pressure due to the attractions in a LJ
fluid displays this $a\rho^2$ dependence even at high
densities~\cite{LJIntPress}. Water, on the other hand, displays a
negative dependence of $P_i$ on density. It is this anomalous behavior
that we seek to explain.

We begin by partitioning the internal energy of the system as
\begin{equation} 
\varepsilon = \varepsilon ^{LJ}+ \varepsilon^{q},
\end{equation} 
where $\varepsilon ^{LJ}$ is the Lennard-Jones contribution to the
energy and $\varepsilon ^{q}$ is the energy due to charge-charge
interactions (note that the change in kinetic energy when perturbing
the volume at constant $T$ is zero, so we only consider the potential
energy). We can then write the internal pressure as
\begin{equation}
P_i=\para{\frac{\partial \varepsilon}{\partial v}}_T=P_i^{LJ}+P_i^{q}.
\label{eq:intpress}
\end{equation}
This decomposition of $P_i$ will allow us to determine which molecular interactions are responsible for the 
strange dependence of this quantity on $\rho$.

\begin{figure}
\centering
\includegraphics[width=0.48\textwidth]{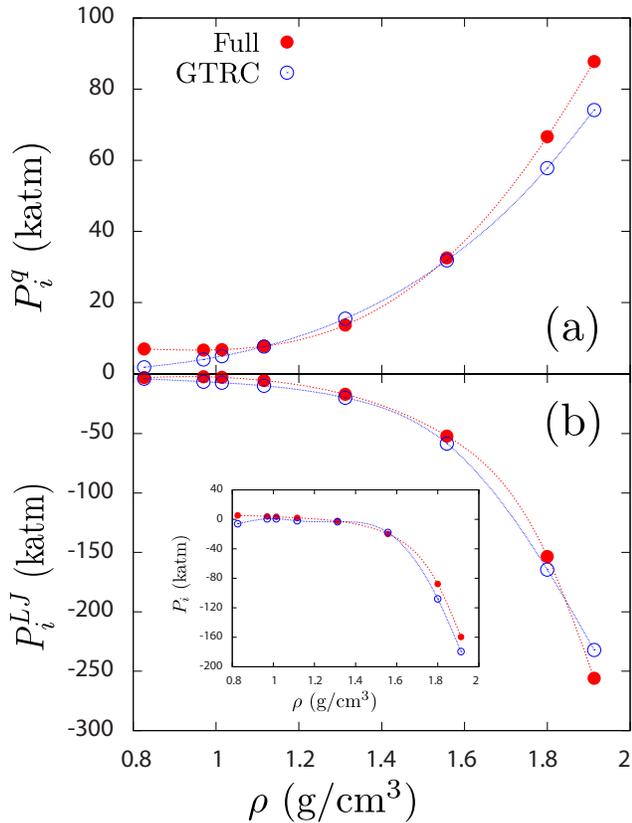}
\caption{(Color online) (a) The electrostatic contribution to the internal pressure,
  $P_i^q$, and (b) the analogous contribution from LJ interactions,
  $P_i^{LJ}$. The total internal pressure as a function of density is
  shown in the inset. Lines are guides to the eye.}
\label{fig:intpress}
\end{figure}

Fig.~\ref{fig:hbondoverlap} suggests the following qualitative
picture.  At a given temperature and density the dominant hydrogen
bond contribution to the energy $\varepsilon$ is determined from the
balance between strong repulsive forces from the LJ cores and strong
attractions from the more slowly varying Coulomb interactions between
donor and acceptor charges.  The Coulomb contribution $P_i^{q}$ to the
internal pressure $P_{i}(T, \rho)$ is positive since a small positive
change in volume reduces the negative Coulomb energy and similarly the
LJ core contribution to $P_i^{LJ}$ is negative. If the density is now
varied at constant temperature we would expect the changes in
$P_{i}(T, \rho)$ to be dominated by the rapidly varying LJ core
forces.  

Conversely, to the extent that the repulsive LJ
cores are like hard spheres, they would contribute no temperature dependence
to the internal pressure at fixed density. 
Thus we expect the more slowly varying Coulomb forces to largely
determine how the internal pressure varies with temperature
at fixed density. The results given below are in complete agreement with these
expectations.

We evaluated Eq.~(\ref{eq:intpress}) by performing MD simulations of
water in the canonical ensemble for various volumes at $T=300$ K.  The
dependence of the internal pressure on density at $T=300$ K is shown
in Fig.~\ref{fig:intpress}. Note that the total internal pressure,
$P_i$, becomes increasingly negative as $\rho$ is increased, in direct
opposition to the $a\rho^2$ dependence given by the vdW equation of
state. However, it is known that as the density of a LJ fluid is
increased to high values so that neighboring repulsive cores begin to
overlap, the total $P_i$ exhibits a maximum, after which the internal
pressure becomes increasingly negative from the dominant contribution
of the repulsive interactions~\cite{LJIntPress}.

As shown in Fig.~\ref{fig:hbondoverlap} there is substantial overlap
of the repulsive LJ cores between nearest neighbors in SPC/E water.
The repulsive interactions from these LJ cores dominate the density
dependence of both $\varepsilon $ and $P_i$ for SPC/E and related
water models, as evidenced by the similarity of the internal pressures
of both the full and GTRC water models in
Fig.~\ref{fig:intpress}. Although $\varepsilon^q> \varepsilon^{LJ}$
for all density, $\varepsilon^q$ does not exhibit very large changes
upon increasing density, a direct consequence of the ability of water
to maintain its hydrogen bond network under the conditions
studied. Thus the density dependence of the internal pressure of SPC/E
water is actually similar to that of a LJ fluid but one at a very high
effective density with substantial overlap of neighboring cores.

\begin{figure}
\centering
\includegraphics[width=0.48\textwidth]{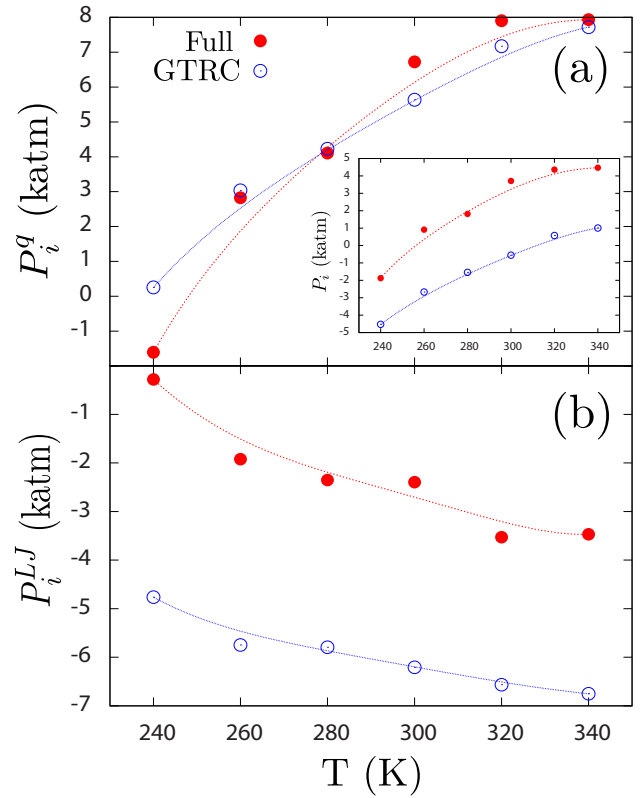}
\caption{(Color online) (a) The electrostatic contribution to the internal pressure,
  $P_i^q$, and (b) the analogous contribution from LJ interactions,
  $P_i^{LJ}$, both as a function of temperature at a fixed volume of
  $v=29.9$~\AA$^3$. The total internal pressure as a function of
  temperature is shown in the inset. Lines are guides to the eye.}
\label{fig:intpressT}
\end{figure}

In addition to the anomalous density dependence of $P_i$, the
temperature dependence of the internal pressure of water has also been
called an anomaly~\cite{PrattWaterQCT-JCP2007}. For most organic
liquids (and vdW fluids), the internal pressure decreases with
increasing temperature, but that of water increases when the
temperature is increased, as shown in Fig.~\ref{fig:intpressT}.  Using
the concepts presented above, we can rationalize this behavior in
terms of molecular interactions.  By decomposing $P_i$ into its
electrostatic and LJ components, we find that $P_i^q$ dominates the
temperature dependence of the internal pressure, increasing with
increasing temperature, while $P^{LJ}_i$ is dominated by repulsive
interactions at all temperatures studied, as evidenced by its negative
value for all $T$. As the temperature of the system is increased, the
number of ideally tetrahedrally coordinated water molecules decreases,
and the hydrogen bond network becomes increasingly
``flexible''. Therefore, if one increases the volume of the system at
high $T$, water will more readily expand to fill that volume. But an
increase in the electrostatic energy will also occur due to a slight
decrease in the number of (favorable) hydrogen bonding
interactions. This will happen to a lesser extent at low temperatures,
when the hydrogen bond network is more rigid and the thermal
expansivity of water is lower.

\section{Unbalanced forces in nonuniform aqueous media from the viewpoint of LMF theory}

\begin{figure}
\centering
\includegraphics[width=0.49\textwidth]{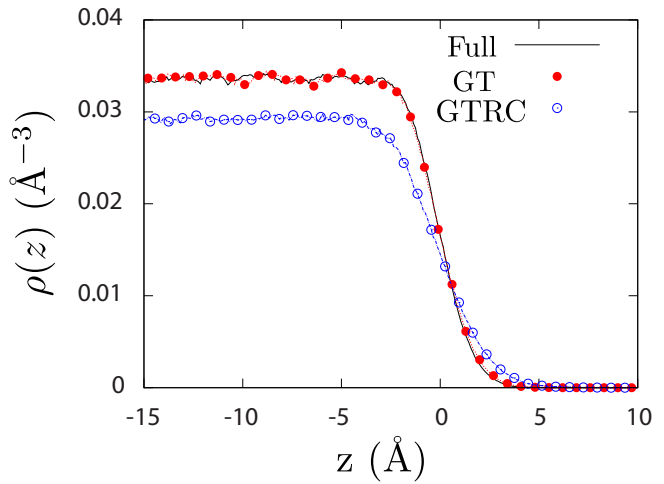}
\caption{(Color online) Density profiles of oxygen sites at the liquid-vapor
  interface of Full, GT, and GTRC water models. The Gibbs dividing
  surface of each interface is located at $z=0$.}
\label{fig:lvint}
\end{figure}

In contrast to uniform systems, a net cancellation of long ranged
forces does not occur in nonuniform environments, and these unbalanced
forces can cause significant changes in the structure and
thermodynamics of the system~\cite{LMFWater,WeeksAnnRevPhysChem}.  As
shown above, the bulk structure of both the GT and GTRC models are
very similar to that of the full water model at a given temperature
and density. But interfacial structure and coexistence thermodynamic
properties of the uncorrected reference models can be very
different. For example, GTRC water still has a self-maintained
liquid-vapor (LV) interface at T = 300 K as illustrated in
Fig.~\ref{fig:lvint}, even though the LJ attractions are ignored,
because of the strong charge pairing leading to hydrogen bond
formation.  However its 90-10\% interfacial width increases to
$w\approx 4.9$~\AA \ from the $w\approx 3.5$~\AA\ seen in both GT
water and the full water model, and the coexisting liquid density of
GTRC water is about about 15\% lower. In contrast, the density profile
of the GT model with LJ interactions fully accounted for is in very
good qualitative agreement with that of the full model. This strongly
suggests that if local hydrogen bonding is properly taken into
account, the equilibrium structure of the LV interface of water is
governed mainly by long ranged LJ attractions, with long ranged
dipole-dipole interactions playing a much smaller role. It is the
exact balance of these long ranged interactions we seek to examine in
this section.

LMF theory provides a framework in which the averaged effects of long
ranged forces are accounted for by an effective external
field~\cite{LMFDeriv}.  It has previously been used mainly as a
computational tool to permit very accurate determination of properties
of the full nonuniform system while using a numerical simulation of
the short ranged reference system in the presence of the effective
field~\cite{LMFOGJPCB,LMFsimp,LMFWater}.  But the effective or
renormalized field also gives a convenient and natural measure of the
importance of long ranged forces in different environments.  In this
section we use the renormalized external fields determined directly
from simulations of interfaces in the full SPC/E water model along
with simulations of truncated water models to quantitatively examine
the relative influence of the local hydrogen bond network and
unbalanced long-ranged Coulomb and van der Waals forces.

\subsection{Water-vapor and water-solid interfaces}

We first consider the LV interfaces of the SPC/E, GT, and GTRC water
models shown in Fig.~\ref{fig:lvint}. The removal of long-ranged
electrostatics in the GT model leaves the density distribution
virtually unchanged, whereas removal of the LJ attractions in GTRC
water has a substantial impact on $\rho(z)$.  To understand this
behavior, we focus our attention on the impact of the averaged
unbalanced forces from the long-ranged electrostatic and LJ
interactions, as determined in LMF theory from the effective external
fields $\Vrl$ and $\phiRlLJ$, respectively and defined below.  The
unbalanced force $\Fr$ acting on an oxygen site from the LMF
potentials is given by
\begin{equation}
\Fr_{\rm O}(\rb)=-\gradr\phiRlLJ(\rb)-q_{\rm O}\gradr\Vrl(\rb).
\end{equation}
Here $q_{\rm O}$ is the partial charge on the oxygen site and
$\Vrl(\rb)$ is the slowly-varying part of the effective electrostatic
field, given by
\begin{equation}
\Vrl(\rb)=\frac{1}{\epsilon}\int d\rb' \rhoq(\rb') v_1\para{\vdiff{\rb}{\rb'}},\label{eq:coulLMF}
\end{equation}
where $\rhoq(\rb)$ is the total charge density of the system.  The
other contribution $\phiRlLJ(\rb)$ is the field arising from the
unbalanced LJ attractions on the oxygen site (where the LJ core is
centered), given by
\begin{equation}
\phiRlLJ(\rb)=\int d\rb' \brac{\rho(\rb')-\rhob}u_1\para{\vdiff{\rb}{\rb'}},
\label{eq:LJLMF}
\end{equation}
with $\rho(\rb)$ indicating the nonuniform singlet density
distribution of oxygen sites and $\rhob$ defined as the bulk density
of oxygen sites at the state point of
interest~\cite{LMFDeriv,WeeksAnnRevPhysChem}.  Since the hydrogen
sites lack LJ interactions, the unbalanced LMF force acting on a
hydrogen site is due exclusively to electrostatics,
\begin{equation}\Fr_{\rm H}(\rb)=-q_{\rm H}\gradr\Vrl(\rb).\end{equation}

\begin{figure}
\centering
\includegraphics[width=0.49\textwidth]{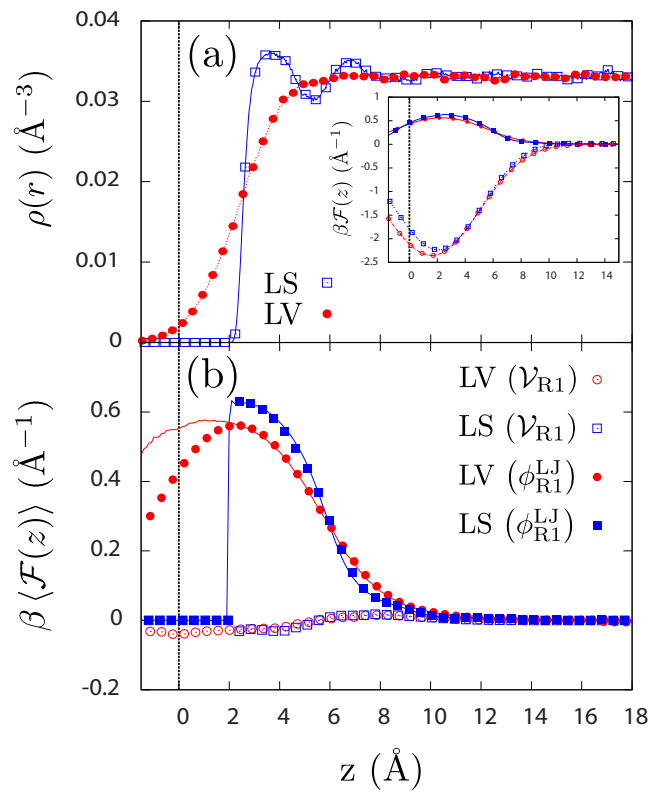}
\caption{(Color online) (a) Density distributions as a function of the $z$-coordinate
  for the hydrophobic LS interface and the LV interface of water. (b)
  Ensemble averaged net force on a water $molecule$ due to $\Vrl$
  (open symbols) and $\phiRlLJ$ (closed symbols) at the LV (circles)
  and LS (squares) interfaces and the density profiles of the two
  systems. Solid lines indicate the net force due to $u_1$. The black
  dashed line at $z=0$~\AA\ indicates the position of the hydrophobic
  wall. The Gibbs dividing interface of the LV system is located at
  $z=2.34$~\AA, in order to make comparison with the water-wall
  interface. $Inset$: Forces on oxygen sites only, determined by
  evaluating the gradient of the corresponding LMF
  potentials. Labeling for the inset is the same as that in (b).}
\label{fig:forces}
\end{figure}

Given its importance in the density distribution of water, it may seem
natural to examine the components of the LMF force on the oxygen
sites, $\Fr_{O}(z)$, shown in the inset of
Fig.~\ref{fig:forces}a. Naive examination of the relative magnitude of
these force functions would lead to the conclusion that long-ranged
electrostatics are the dominant unbalanced force at the LV
interface. However, $\Vrl$ also interacts with hydrogen sites and one
should instead consider the net forces from long ranged Coulomb and LJ
interaction felt by an entire water \textit{molecule} at these
interfaces.

This ensemble averaged net molecular force $\avg{\Fr}$
(Fig.~\ref{fig:forces}b) clearly indicates that the net unbalanced
force at an interface is almost entirely due to long-ranged LJ
attractions from the bulk, which pull water molecules away from the
interface. The long-ranged Coulomb contributions to the average force
on a water molecule are essentially negligible in comparison. This is
not surprising since water molecules are neutral, and it has
previously been shown that the small net interfacial electrostatic
force simply provides a slight torque on water molecules in this
region~\cite{LMFWater}.  This torque has little effect on the oxygen
density distribution, as illustrated by the good agreement of the GT
model density profile with that of the full water model in
Fig.~\ref{fig:lvint}.  However, it plays a key role in determining
electrostatic and dielectric properties, which are strongly affected
by the behavior of the total charge density, and here the uncorrected
GT model gives very poor results~\cite{LMFWater,LMFDeriv}.

It is also instructive to compare the unbalanced long ranged forces at
the LV interface to those at the liquid-solid (LS) interface between
water and a model hydrophobic 9-3 LJ wall introduced by Rossky and
coworkers~\cite{LeeMR}, as shown in
Fig.~\ref{fig:forces}b. Despite the large differences in the density
profiles shown in Fig.~\ref{fig:forces}a, the net unbalanced forces
on molecules at the LV and LS interfaces are remarkably similar for
all $z$ until molecules encounter the harsh repulsion of the wall and
an accurate sampling of $\avg{\Fr(z)}$ by simulation cannot be
made. Water molecules can sample all regions in the liquid-vapor
interface, leading to a smooth $\avg{\Fr(z)}$ at smaller $z$. 

  Indeed, the net molecular force due explicitly to a configurational
  average of the attractive $u_1(r)$ acting on molecules present at
  each $z$-position is in outstanding quantitative agreement with that
  arising from $\phiRlLJ(z)$ for all adequately sampled regions in the
  liquid, as illustrated by the solid lines in Fig.~\ref{fig:forces}b.
  This serves largely as confirmation of the validity of the
  mean-field treatment inherent in LMF theory within the liquid
  slabs. Deviations between the two quantities for distances less
  than the Gibbs dividing surface are a reflection of the increasing
  effect of larger force fluctuations due to long-wavelength capillary waves
  not well described by mean field theory.
The relative magnitudes of the components of $\avg{\Fr}$ for the LV
and LS interfaces are strikingly similar, with the net unbalanced LJ
force $\avg{\Fr\para{z;\phiRlLJ}}$ reaching its maximum value of
slightly less than $\kT/$\AA\ near the Gibbs dividing interface and
the repulsive boundary of the wall, respectively.

The similarities of the unbalanced forces at the LV and the
hydrophobic LS interfaces of water and the dominance of the LJ
attractions are completely consistent with the analogies commonly
drawn between these two
systems~\cite{StillingerLV,ChandlerNatureReview,DewettingRev} and used
in the LCW theory of
hydrophobicity~\cite{LCW,VarillyHphobLattice,WeeksAnnRevPhysChem}.  A
common criticism of LCW theory is its apparent neglect of the hydrogen
bond network of water and the use of a van der Waals like expression
for the unbalanced force at an interface. Although some features of the network are
implicitly captured by using the experimental surface tension and
radial distribution function of water as input to the theory,
electrostatic effects at the interface, including dipole-dipole
interactions, are ignored.  However, this assumption is justified
since the averaged effects of long-ranged dipole-dipole interactions,
accounted for by $\Vrl$, are shown to indeed be negligible at a
hydrophobic interface (Fig.~\ref{fig:forces}).  LCW theory correctly
describes the unbalanced LJ attractions from the bulk, which dominates
the behavior at both the liquid-vapor and extended hydrophobic
interfaces.

\subsection{Crossover from small to large length scale hydration}

LCW theory combined the idea of unbalanced forces with experimental
data in water to predict a crossover in the solvation of spherical
hydrophobic solutes occurring at a radius of about 1~nm~\cite{LCW}.
While the local hydrogen bond network can be maintained around smaller
solutes, some bonds must be broken on larger length scales, leading to
an enthalpically dominated
regime~\cite{StillingerLV,ChandlerNatureReview,DewettingRev}.  Here we
use LMF theory and a direct analysis of the unbalanced long ranged
forces to provide further physical insight into how this crossover
comes about. We study the hydration of spherical hydrophobic solutes,
which interact with the oxygen site of water via a solute-water LJ
potential $u_{\rm sw}(r)$ with a fixed well depth of $\varepsilon_{\rm
  sw}=0.19279$ kcal mol$^{-1}$ and varying solute-water interaction
length-scales, $\sigma_{\rm sw}$, ranging from 2.0~\AA\ to
15.0~\AA. This size range spans the crossover from small to large
length-scale hydrophobic hydration as determined by both simulation
and theory~\cite{AshbaughCrossover,LCW}.

\begin{figure}
\centering
\includegraphics[width=0.49\textwidth]{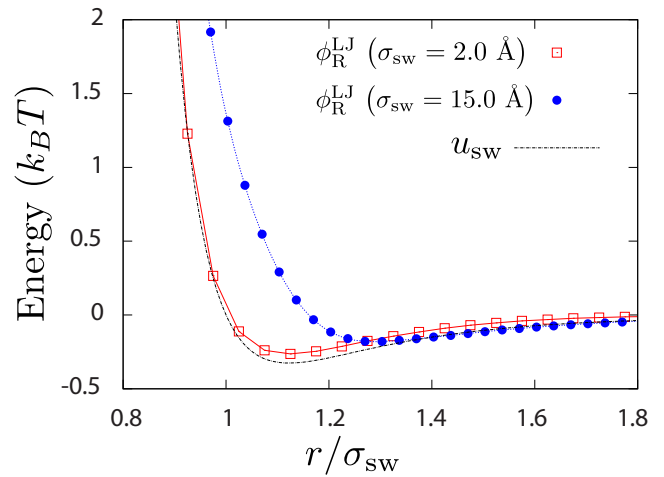}
\caption{(Color online) Comparison of renormalized solute-water vdW potentials for
  $\sigma_{\rm sw}=2.0$~\AA\ (squares) and $\sigma_{\rm sw}=15.0$~\AA\
  (circles). The full solute-water LJ potential $u_{\rm sw}$ is also
  shown for comparison (black dash-dot line). Note that the x-axis is
  scaled by the solute-water interaction length-scale $\sigma_{\rm
    sw}$.}
\label{fig:compots}
\end{figure}

As shown earlier, the unbalancing force from LJ attractions dominates
that due to long ranged electrostatics at nonpolar interfaces so we
focus only on the LJ forces in our analysis of the length scales of
hydration. The total renormalized external solute-solvent field is the
sum of the bare solute-solvent field and the slowly-varying LMF
potential $\phiRlLJ(r)$ in Eq.~(\ref{eq:LJLMF}), which accounts for
unbalanced LJ forces from the nonuniform water (oxygen) density
distribution around the solute. It can be written as
\begin{equation}
\phiRLJ(r)=u_{\rm sw}(r)+\phiRlLJ(r).
\end{equation}
We compare these potentials for small and large LJ solutes in
Fig.~\ref{fig:compots}.  At small solute sizes, the renormalized field
$\phiRLJ$ exhibits a repulsive core nearly identical to that of the bare
$u_{\rm sw}$, and the effective attractions are hardly altered upon
renormalization of the potential, indicating only a small unbalanced
force around a small solute. The drive to maintain the hydrogen bond
network around small solutes dominates the water structure, and
solute-water and water-water LJ attractions are found to have little
effect on the solvation structure~\cite{ChandlerNatureReview}.

In contrast, water cannot completely preserve its hydrogen bond
network at an interface around a large solute and one hydrogen bond
per interfacial water molecule tends to be broken on
average~\cite{StillingerLV,LeeMR,ChandlerNatureReview,DewettingRev}.
This creates a soft fluctuating interface for which partial drying can be
induced by unbalanced attractions from the bulk. These are
re-expressed in LMF theory as an effective repulsion pushing water
away from the solute surface, as illustrated by the renormalized
potential for the large solute in Fig.~\ref{fig:compots}. In order for
water to wet such an extended solute surface, this large effective
repulsion must be overcome.  In hydrophilic surfaces this typically
arises from strongly attractive polar groups on the surface, such as
hydrogen bonding sites or charged groups, which also can strongly
perturb local hydrogen bond configurations.

\begin{figure*}
\centering
\includegraphics[width=0.95\textwidth]{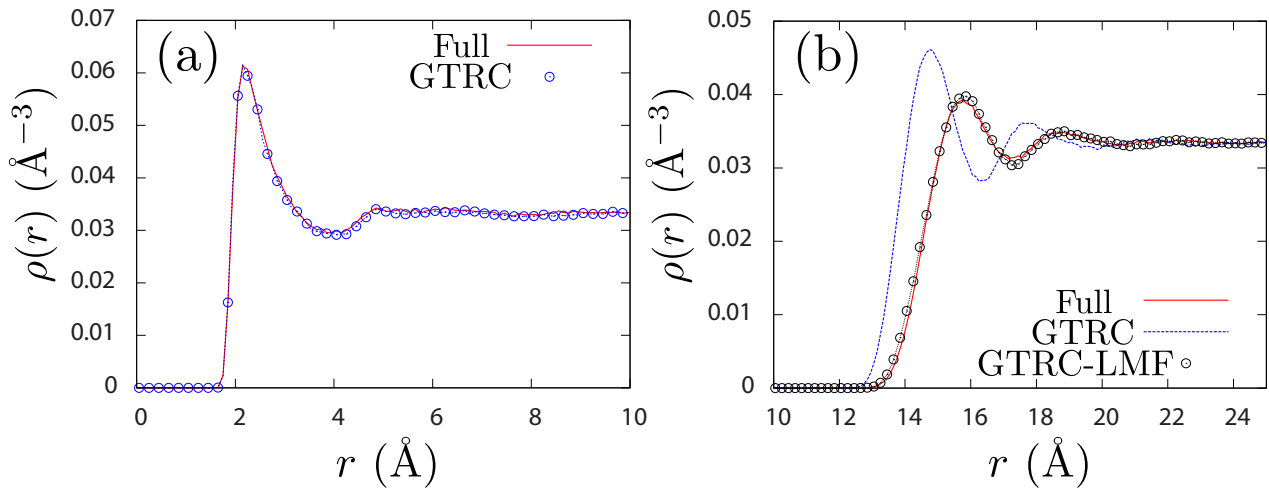}
\caption{(Color online) Singlet density distributions of water oxygen sites,
  $\rho(r)$, around LJ solutes with $\sigma_{\rm sw}=2.0$~\AA\ (a) and
  $\sigma_{\rm sw}=15.0$~\AA\ (b) for Full and GTRC water models at 
  pressures of $P=1.0$~atm and $P_0=P-P_1$, respectively. Data
  for GTRC water in the presence of the renormalized LMF solute
  potential is also shown for the large solute (GTRC-LMF).}
\label{fig:rhor}
\end{figure*}

To illustrate the interplay between unbalanced forces from the bulk,
water-water hydrogen-bonding, and the length-scale dependence of
hydration, we study the hydration of a small ($\sigma_{\rm
  sw}=2.0$~\AA) and large ($\sigma_{\rm sw}=15.0$~\AA) LJ solute by
SPC/E water and the short-ranged GTRC water model.  Since the GTRC
model lacks LJ attractions, the unbalancing force from the bulk is
absent but the local hydrogen bonding network of water is accurately
captured. The GTRC model thus properly describes the dominant
crossover behavior of retaining or breaking hydrogen bonds in the
small or large length scale regimes, respectively, but the subsequent
large length scale interface properties will be incorrect.  Comparing
GTRC water with the full water model provides more insight into the
relative importance of the local network and the unbalancing force on
the structuring of water around each solute.

Singlet density distributions of the water oxygen sites around the
small and large LJ solute are shown in Fig.~\ref{fig:rhor}a
and~\ref{fig:rhor}b, respectively. In the small solute regime,
$\rho(r)$ for GTRC water is essentially identical to that of the full
water model.  This provides dramatic confirmation of the standard
physical picture that small length scale hydrophobicity is almost
completely dominated by the need to maintain local hydrogen bonding
around the solute. Unbalanced forces from either the LJ or long-ranged
Coulomb interaction play essentially no structural role in this
regime.

In contrast, although the GTRC model correctly describes the necessary
breaking of local hydrogen bonds around a large solute in
Fig.~\ref{fig:rhor}b the details of the resulting interface profile
are very different from that of the full water model. Removal of LJ
attractions in GTRC water eliminates the large effective repulsion at
the extended solute surface, and uncorrected GTRC water appears to wet
the surface of the nonpolar solute. However, the interface is soft
and when the unbalanced force
from $\phiRlLJ$ in Fig.~\ref{fig:forces} as given by LMF theory is
also taken into account, depletion and the correct structuring of
water at the solute surface is very accurately described by the
GTRC-LMF curve in Fig.~\ref{fig:rhor}b.

Although the qualitative dependence of the unbalanced force on
solvation length scale is depicted in Fig.~\ref{fig:compots}, a more
quantitative metric of this behavior is desirable. To that end, we
introduce the mean solvation force acting on water due to the
renormalized LJ solute external field,
\begin{eqnarray}
  F_S\brac{\phi^{\rm LJ}_{\rm R}}&=&-\int d\rb \rho(r)\frac{\partial u_{\rm sw}(r)}{\partial r}-\int d\rb \rho(r)\frac{\partial \phiRlLJ(r)}{\partial r}\nonumber \\
  &=& F_S\brac{u_{\rm sw}} + F_S\brac{\phiRlLJ}.
\label{eq:solvforce}
\end{eqnarray}

\begin{figure}
\centering
\includegraphics[width=0.49\textwidth]{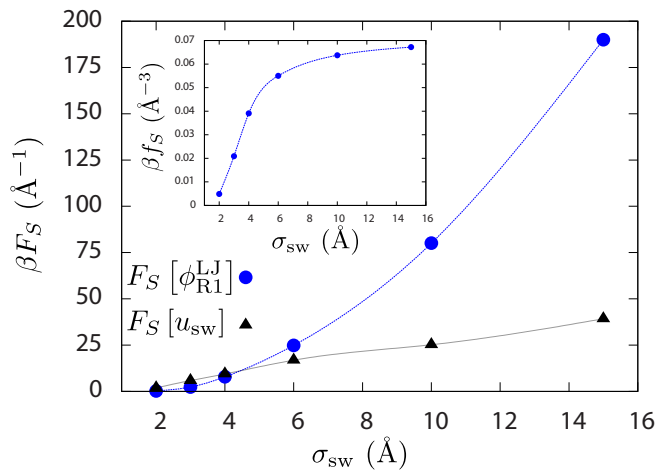}
\caption{(Color online) Components of the mean solvation force $F_S$ due to
  $\phiRlLJ$ and $u_{\rm sw}$ as a function of the solute-water
  interaction length scale, $\sigma_{\rm sw}$. The inset displays the
  solvation force due to $\phiRlLJ$ scaled by the surface area of the
  solute.}
\label{fig:solvforce}
\end{figure}

As illustrated in Fig.~\ref{fig:solvforce}, the unbalanced force due
to water-water LJ attractions dominates in the large-scale regime, and
can only be overcome by unphysically large solute-water LJ
attractions. In the small scale regime, solute-water attractions are
comparable in magnitude to the unbalancing force, and can be larger
for certain solute sizes (the relative magnitudes are dependent upon
the value of $\varepsilon_{\rm sw}$). However, as the GTRC model
shows, the important physics in this regime is simply maintaining the
hydrogen bond network, and this imposes a near constant solvation
structure as the water-solute attractions are
varied~\cite{HuangChandlerSoluteSolvAttr,ChandlerNatureReview}.

Upon normalizing the mean unbalanced force by the surface area of the
solute,
\begin{equation}
  f_S\brac{\phiRlLJ}=F_S\brac{\phiRlLJ}/4\pi\sigma_{\rm sw}^2,
\end{equation} 
a transition from scaling of $F_S$ with solute volume to scaling with
surface area occurs at the small to large length scale solvation
crossover length of roughly 10~\AA, as evidenced by the plateau in
$f_S$ depicted in the inset of Fig.~\ref{fig:solvforce}. This plateau
is similar to that which occurs in the solvation free
energy~\cite{LCW,scalingHphobFreeEnergies}, and provides another indicator
of the transition from small to large length scale hydrophobicity.

\section{Conclusions}

In this work, we have examined the different roles of short and long
ranged forces in the determination of the structure and thermodynamics
of uniform and nonuniform aqueous systems, using concepts inherent in
classical perturbation and LMF theory.  In particular, we have
evaluated individually for SPC/E water the contributions of (i) all
the strong short ranged repulsive and attractive interactions that
lead to the local hydrogen-bonding network, (ii) longer ranged
dispersive LJ attractions between molecules, and (iii) long ranged
dipole-dipole interactions, and demonstrated a hierarchical ordering
of their importance in determining several properties of water in
uniform and nonuniform systems.
  
All of our truncated models accurately describe the local hydrogen
bonding network, and as expected, this network alone is sufficient to
match bulk structure as well as solvent structure around small
hydrophobic solutes provided that the bulk density and temperature are
accurately prescribed.  Furthermore, the anomalous temperature and
density dependence of the ``internal pressure'' of water is found to
be dominated by the competing short-ranged repulsive and attractive
forces determining the local hydrogen bonding network as well.
  
But local network concepts alone cannot capture all the complexities
of even the simple SPC/E water model.  While the dispersive LJ attractions
between water molecules primarily provide a uniform cohesive energy in
bulk systems, they strongly influence the structure and density
profile of large scale hydrophobic interfaces.  Their importance
provides further support for analogies between water at extended
hydrophobic interfaces and the liquid-vapor interface, and the
unbalanced LJ force can be used to quantify the transition between
small and large scale hydrophobicity for simple solutes.

Although the long-ranged dipolar interactions between molecules have
only small effects on most of the interfacial density properties
considered here, we have shown elsewhere that they are crucial in
determining dielectric properties of both bulk and nonuniform
water. Indeed, as will be discussed elsewhere, we have found that
electrostatic quantities may in fact be a sensitive structural probe
of hydrophobicity in general
environments~\cite{WaterElectrostaticsHphobProbe}.

This interaction hierarchy, wherein strong short-ranged local
interactions alone determine structure in uniform environments while
the longer ranged forces are needed as well to capture other
properties could prove quite useful in refining simple site-site water
models. Current water models incorporate a vast amount of clever
engineering and empirical fine-tuning and manage to reproduce a
variety of different properties through a complex balance of competing
interactions with simple functional forms.  Changes in the potential
that improve one property generally speaking produce poorer results
for several others.

One promising route to a more systematic procedure may be sensitivity
analysis, in which small perturbations of potential parameters are
made and the correlated response of a variety of physical observables is
quantified.  By perturbing the relative magnitudes of short and long
ranged interactions, Iordanov \etal \ found that thermodynamic
properties of bulk water are most sensitive to small changes in the LJ
repulsions and the short ranged electrostatic
interactions~\cite{SensitivityAnalysisDecomp}, in agreement with our
findings.  A new water model was then proposed by optimizing
parameters to reproduce a specific bulk thermodynamic quantity (the
internal energy) in an attempt to correct the deficiencies present in
a previously developed water potential.
 
However, the theoretical scheme of splitting the potential described
in this paper may provide a more concrete and physically suggestive
path to incrementally match various known physical quantities for
water without ruining the fitting of previous quantities, and one
could combine an approach like sensitivity analysis with the
conceptual framework presented herein to systematically optimize a
specific water model.
  
In particular, it has recently been suggested that the accuracy with
which a water model can predict the experimental $T_{MD}$ correlates
well with the accuracy that the same model displays in predicting the
thermodynamics of small-scale hydrophobic hydration~\cite{AshbaughTMD}. Arguably,
the least justified feature of current simple water models like SPC/E
is the functional form of the core LJ potential $u_{0}(r)$, especially
at the very short separations relevant for describing local hydrogen
bonding as illustrated in Fig.~\ref{fig:hbondoverlap}.  One could try
to fine-tune a GTRC-type model through alteration of the local
hydrogen bond network by varying the form of the repulsive core in
order to match the experimental density maximum, as well as other bulk
properties like the internal pressure, in order to obtain a
short-ranged system that yields accurate bulk properties.  Although a
detailed discussion of this process is beyond the scope of this
article, one could try to use some type of optimization procedure to
determine such
potentials~\cite{TorquatoPotentialOpt,SAWaterThermo93,SensitivityAnalysisDecomp}.
Perhaps first principles DFT simulations
\cite{DFTwaterLV,DFTNPTwater} could be used to provide a more fundamental
description of the local network. Subsequently, the structure and thermodynamics of nonuniform systems,
which require dispersions and long ranged Coulomb interactions, could
be used to parametrize the long-ranged interactions.
 
 \begin{acknowledgments}
This work was supported by the National Science Foundation (grants
CHE0628178 and CHE0848574). We are grateful to Lawrence Pratt and
Shule Liu for helpful remarks. We also thank an anonymous reviewer
for bringing references~\cite{DFTwaterLV} and~\cite{DFTNPTwater} to our
attention.
\end{acknowledgments}
\begin{table}[!t] 
\centering
\begin{tabular}{c c c c c}  
\hline\hline                        
System & $N$ & $T$ & $t_{equil}$ & $t_{run}$  \\ [0.5ex] 
\hline                    
Bulk Water & 256 & 220-340 K & 5 ns & 5 ns  \\   
Bulk Water & 1000 & 220-300 K & 2 ns & 2 ns  \\
LV Interface & 1728 & 298 K & 1 ns & 500 ps  \\
LS Interface & 2468 & 298 K & 1 ns & 500 ps  \\
Small LJ Solutes ($\sigma_{\rm sw}<10$ \AA) & 1000 & 298 K & 2 ns & 2 ns  \\
Large LJ Solutes ($\sigma_{\rm sw}\ge10$ \AA) & 6000 & 298 K & 1 ns & 1 ns  \\[1.0ex]
\hline     
\hline
\end{tabular}
\caption{Details of the MD simulations performed in this work. $N$ and $T$ refer to the number of water molecules and the temperature of the system, respectively. The equilibration times and data collection times are denoted by $t_{equil}$ and $t_{run}$, respectively.} 
\label{table:Sim}  
\end{table} 

\appendix*
\section{Simulation Details}

All molecular dynamics simulations were performed using modified
versions of the DL\_POLY software package~\cite{dlpoly} and the SPC/E
water model~\cite{SPCE} or its variants described in Section 2. The
equations of motion were integrated using the leapfrog algorithm with
a timestep of 1~fs~\cite{CompSimLiqs} while maintaining constant
temperature and pressure conditions through the use of a Berendsen
thermostat and barostat respectively~\cite{BerendsenBaroThermo}.

\subsection{Bulk water simulations}

The evaluation of electrostatic interactions in bulk simulations of
the full SPC/E water model employed the standard Ewald summation
method using a real space cutoff of 9.5~\AA, unless this was larger
than half of the box length, in which case the cutoff was set to half
of the box length~\cite{CompSimLiqs}.  Short-ranged electrostatic
interactions in the GT and GTRC reference systems, as well as LJ
interactions in all systems, were truncated at the real space cutoff
length used in the analogous full system. Simulations of bulk water
were performed with $N=1000$ molecules in the isothermal-isobaric
(NPT) ensemble to determine the density maximum and with $N=256$
molecules in the canonical (NVT) ensemble to determine $P(T)$ and the
internal pressure. The duration of equilibration and production runs,
as well as the temperatures sampled are listed in Table 1. The
internal pressure in Eq.~(\ref{eq:intpress}) was calculated by
evaluating $\varepsilon(v)$ for numerous values of $v$ at each
$T$. The function $\varepsilon(v)$ was then fit to a polynomial, which
was differentiated at the desired $v$ to yield the internal pressure.

\subsection{Simulation of nonuniform systems}

In order to generate starting configurations for the LV and LS
interfacial systems discussed in Section 5, we first equilibrated $N$
water molecules in a cubic geometry, where $N$ is listed in Table
1. The $z$-dimension of the system was then elongated to more than
three times the $x$- and $y$-dimensions, and in the case of the LS
interface, a wall potential of the form
\begin{equation}
U_w(z)=\frac{A}{\len{z-z_w}^9}-\frac{B}{\len{z-z_w}^3}
\end{equation}
was added at $z_w=0$ and the parameters $A$ and $B$ are given in
Ref.~\cite{RosskyJCP}. In order to ensure water molecules did not
approach the wall from $z<0$, a repulsive wall was added at large $z$
to constrain the water molecules to the desired region of the
simulation cell while still allowing a large vacuum region for the
formation of a vapor phase. Electrostatic interactions were handled
using the corrected Ewald summation method for slab
geometries~\cite{Ewald3DC} with a real space cutoff of 11.0~\AA, which
was also the cutoff distance for LJ and short-ranged electrostatic
interactions.

Molecular dynamics simulations of the hydration of LJ solutes were
performed in the NPT ensemble using standard Ewald summation to
evaluate the electrostatic interactions, with a real space cutoff of
11.0~\AA. The short ranged $v_0$ potential and the water-water LJ
potential were also truncated at 11.0~\AA. The LJ solute was
represented by a fixed external potential centered at the origin, and
the solute-water interactions were truncated at one-half the length of
the simulation cell. The number of molecules and simulation times are
listed in Table 1.


\end{document}